\documentclass[reprint, superscriptaddress, secnumarabic, amssymb, nobibnotes, aps, pra]{revtex4-1}

\usepackage{graphicx}
\usepackage{subfiles}
\usepackage{epstopdf}
\usepackage{parskip}
\usepackage[T1]{fontenc}
\usepackage[latin9]{inputenc}
\usepackage{amsbsy}
\usepackage{gensymb}

\usepackage[T1]{fontenc}
\usepackage[latin9]{inputenc}
\usepackage{amsmath}
\usepackage{amssymb}
\usepackage{bbm}
\usepackage{physics}
\usepackage{braket}
\usepackage{xcolor}
\usepackage{ wasysym }
\usepackage{float}
\allowdisplaybreaks
\usepackage{graphicx}
\usepackage[colorlinks=true]{hyperref}  
\hypersetup{
    bookmarks=true,         
    unicode=false,          
    pdftoolbar=true,        
    pdfmenubar=true,        
    pdffitwindow=false,     
    pdfstartview={FitH},    
    pdftitle={Ce$_{2}$Ir$_{3}$Ga$_{5}$ : a new locally non-centrosymmetric heavy fermion system},    
    pdfauthor={Arushi et al.},     
    pdfsubject={},   
    pdfcreator={},   
    pdfproducer={}, 
    pdfkeywords={} {} {}, 
    pdfnewwindow=true,      
    colorlinks=true,       
    linkcolor=blue, 
    citecolor=blue,        
    filecolor=magenta,      
    urlcolor=blue           
} 
\usepackage[normalem]{ulem}

\newcommand{\figref}[1]{Fig.~\ref{#1}}

\renewcommand{\approx}{\simeq}

\makeatletter
\def\maketitle{
\@author@finish
\title@column\titleblock@produce
\suppressfloats[t]}
\makeatother

\begin{document}
\title{Ce$_{2}$Ir$_{3}$Ga$_{5}$ : a new locally non-centrosymmetric heavy fermion system}
\author{Arushi}
\email[Arushi@cpfs.mpg.de]{}
\author{Raul Cardoso-Gil}
\author{Christoph Geibel}
\affiliation{Max-Planck Institute for Chemical Physics of Solids, 01187 Dresden, Germany}
\date{\today}
\begin{abstract}
\begin{flushleft}
\end{flushleft}
Recently, a new type of unconventional superconductivity with a field-induced transition between two different superconducting (SC) states was discovered in the heavy fermion system CeRh$_{2}$As$_{2}$. This unusual SC state was proposed to be based on specific symmetries of the underlying structure, i.e., a globally centrosymmetric layered structure, but where the Ce-layers themselves lack inversion symmetry. This new type of SC state has attracted strong interest, prompting the search for further heavy fermion systems crystallizing in structures with appropriate symmetries. We report the discovery and the study of a new Ce-based heavy fermion system with a globally centrosymmetric structure but without inversion symmetry on the Ce-site, Ce$_{2}$Ir$_{3}$Ga$_{5}$. A single crystal X-ray diffraction study revealed an orthorhombic U$_{2}$Co$_{3}$Si$_{5}$ type structure. Resistivity, specific heat, and magnetization measurements indicate a moderate-heavy fermion behavior with a Kondo energy scale of the order of 40 K. Most experimental results suggest the absence of magnetic order, but a tiny anomaly in the specific heat opens the possibility for a very weak, itinerant type of ordering.

\end{abstract}
\maketitle

\section{Introduction}
In the past 20 years, the absence of inversion symmetry (IS) has emerged as an effective parameter to induce unusual superconducting properties and unconventional superconducting states \cite{NCS_Rev1,NCS_Rev2}. Expected and reported effects are particularly strong in so-called heavy fermion systems \cite{NCS_SCES_1,NCS_SCES_2}. Here, the strong electronic correlation effects emerging from the hybridization between localized $f$ electrons, e.g. from Ce, Yb, or U-atoms, and itinerant valence electrons results in the formation of quasiparticles with huge effective masses at the Fermi level, with profound consequences on electric transport and thermodynamic properties. Among others, the enormous effective masses result in large SC critical fields (H$_{c2}$) of several Teslas despite very low SC transition temperatures (T$_{c}$) of the order of 0.5 K \cite{CePt3Si,CeIrSi3_1}. Absence of IS can then induce further unusual strong effects on these SC properties. Evidence of a strong influence of the absence of IS was first reported for a series of compounds crystallizing in BaNiSn$_{3}$ structure such as CeIrSi$_{3}$, CeRhSi$_{3}$, CeCoGe$_{3}$, and CeIrGe$_{3}$, where superconductivity under pressure was observed to present huge H$_{c2}$'s with a large anisotropy \cite{CeIrSi3_1,CeRhSi3,CeCoGe3,CeIrGe3}. These unusual SC properties are assumed to emerge from the absence of IS in the structure of these compounds. Furthermore, the lack of an inversion symmetry also give rise to unconventional magnetic \cite{UM1,UM2,UM3,UM4} and normal state properties \cite{NMS1}.

Later on, theoretical studies indicated that even in globally centrosymmetric structures, the absence of an inversion symmetry on a local level can induce unusual SC states \cite{LNCS1,LNCS2,LNCS3,LNCS4}. Thus, in layered structures without inversion symmetry in the layers but inversion symmetry for the whole structure, the latter results in an additional quantum number: the parity of the electronic states respective to the inversion center. Theoretical studies demonstrated the possibility of forming an unconventional SC state with odd parity in zero fields, which gets replaced by an even parity SC state under an appropriate magnetic field \cite{LNCS_even-odd_1,LNCS_even-odd_2,LNCS_even-odd_3}. The latter one presents a huge H$_{c2}$ anisotropy. Remarkably, exactly such an unconventional SC phase diagram was recently observed in the heavy Fermion system CeRh$_{2}$As$_{2}$ \cite{CeRh2As2_1}. The nature of the unusual SC and normal state properties of CeRh$_{2}$As$_{2}$ are now the subject of strong interest and intense research \cite{CeRh2As2_2,CeRh2As2_3,CeRh2As2_4,CeRh2As2_5,CeRh2As2_6}.

An obvious question is whether similar unconventional and intriguing properties can be found in other heavy fermion systems with a global inversion symmetric structure but without IS on the Ce site. There are several Ce compounds with appropriate structures, but only some are appropriate for the search for unconventional superconductivity. For a Ce compound to be a promising candidate, its strength of the interaction between the 4$f$ and conduction electrons, i.e., the $f$-$c$ hybridization, has to be fine-tuned to put the system close to the so-called quantum critical point (QCP), where the $f$-$c$ hybridization is just sufficiently strong to destroy the magnetic order, resulting in a transition to a paramagnetic ground state. Heavy Fermion behavior and unconventional superconductivity are only expected and observed in the vicinity of such a QCP. Ce, Yb, or U-compounds with suitable structural properties and appropriate strength of the $f$-$c$ hybridization are very scarce \cite{LNCS4,CeRh2As2_1,UTe2,UCoGe_UPt3}.

In a search for appropriate systems, we explored the ternary Ce-Ir-Ga phase diagram. Combining the Ir and Ga ligands seemed to be a promising approach since Ir promotes a quite strong $f$-$c$ hybridization, while Ga induces a comparatively weak one. This search leads to the discovery of a yet unknown compound, Ce$_{2}$Ir$_{3}$Ga$_{5}$. X-ray diffraction demonstrated that it crystallizes in the U$_{2}$Co$_{3}$Si$_{5}$ structure type, which is globally centrosymmetric, but where the (crystallographic unique) Ce site lacks inversion symmetry. Electrical resistivity, magnetic susceptibility, and specific heat measurements evidenced a sizeable $f$-$c$ hybridization, resulting in a corresponding Kondo energy scale of the order of 40 K. Further properties suggest that Ce$_{2}$Ir$_{3}$Ga$_{5}$ is on the paramagnetic side of the QCP and presents a heavy Fermi liquid ground state. However, a weak anomaly in the specific heat could be an indication of a very weak magnetic order of itinerant 4$f$ electrons. We could not yet observe superconductivity, which could be due to a too-large residual resistivity.

\section{Experimental Details}

For the synthesis of Ce$_{2}$Ir$_{3}$Ga$_{5}$ samples we started from the constituent elements  Ce (99.99\%), Ir (99.99\%) and Ga (99.999\%). Ce and Ir were taken in the stoichiometric ratio, while for Ga, we added 1\% extra to compensate for evaporation losses. A differential thermal analysis (DTA) study indicated a congruent melting at 1340 $\degree$C. Accordingly, we tried two different methods for the preparation of this compound. The first method involves melting the elements in an arc furnace under an argon atmosphere. The resulting ingot was flipped over and remelted, and the same procedure was repeated four to five times to ensure homogeneity. In the second method, we used the pre-reacted ingot obtained by arc melting, crushed it into small pieces, and transferred it into a carbon crucible enclosed in a tantalum cylinder. The tantalum cylinder was then closed using an arc furnace under an argon atmosphere of 800 mbar. The whole crucible was then put in a resistance furnace, and the following heat treatment was applied: ramped up to 1380~$\degree$C, dwelled for 15 minutes, and then slowly cooled to room temperature. Both methods lead to a phase pure Ce$_{2}$Ir$_{3}$Ga$_{5}$ sample. Since, to the best of our knowledge, this compound has not been reported before, its crystallographic structure was determined using both single-crystal and powder X-ray diffraction (XRD). Single-crystal XRD data was obtained from a Rigaku AFC7 diffractometer system (MoK$_{\alpha}$ radiation, $\lambda$ = 0.71073 \AA), whereas powder XRD pattern was collected using Stoe-Stadi-MP powder diffractometer in transmission mode equipped with CuK$_{\alpha1}$ radiation ($\lambda$ = 1.54056 \AA), curved germanium (111) Johann-type monochromator, and DECTRIS MYTHEN2 1K silicon strip detector. The phase purity of the Ce$_{2}$Ir$_{3}$Ga$_{5}$ samples was confirmed by energy dispersive X-ray spectroscopy (EDXS). Magnetization measurements were carried out using Quantum design SQUID in the temperature range 1.8 K - 300 K and up to 7 T external field. AC transport and specific heat measurements were performed in a physical property measurement system (PPMS) equipped with a $^{3}$He probe (0.5 K). Four-probe and two-tau relaxation methods were employed for AC transport and specific heat experiments, respectively.

\section{Results and Discussion}
\subsection{Structural Characterization}

\begin{figure}[t]
\includegraphics[width=1.0\columnwidth,origin=b]{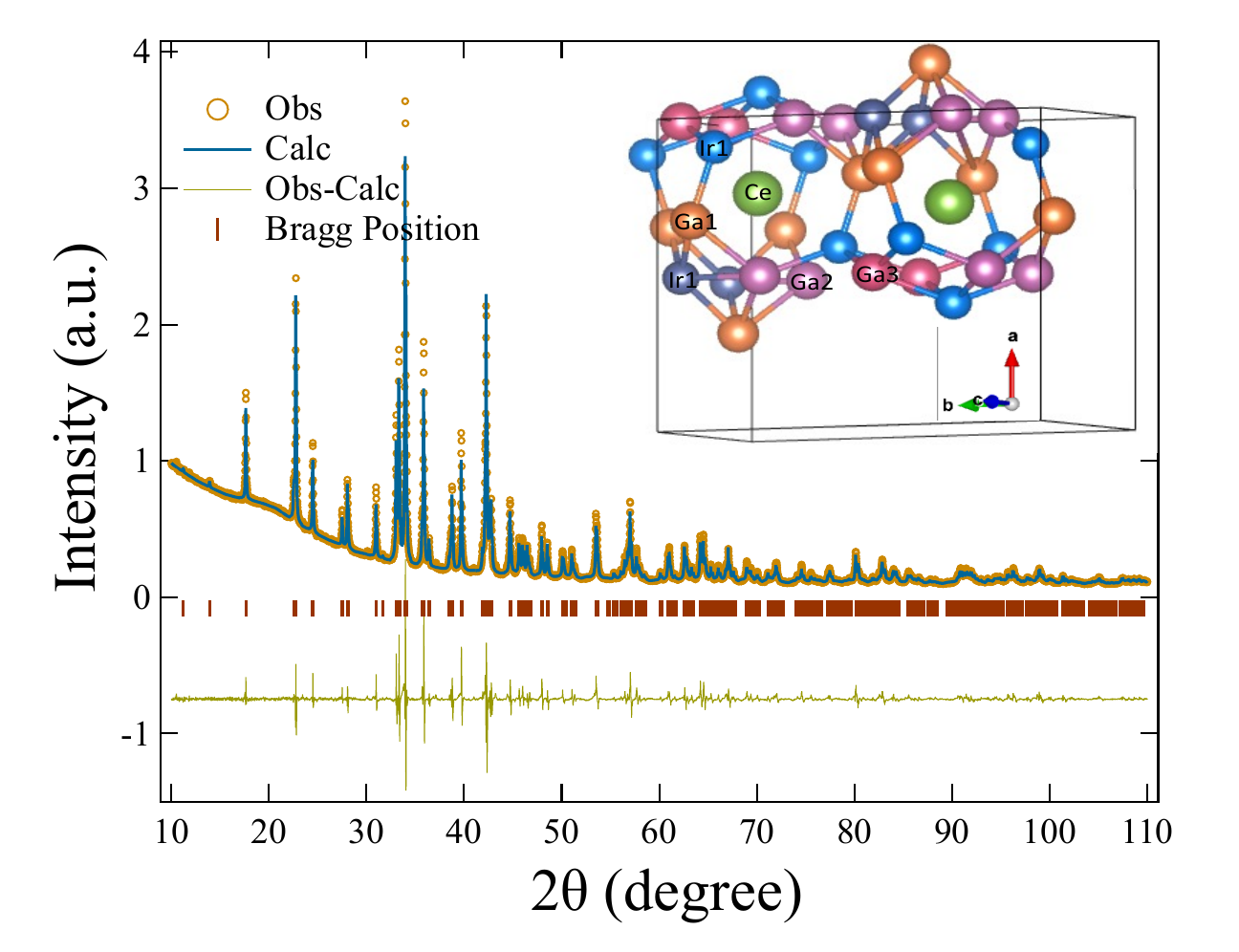}
\caption{\label{Fig1:xrd} Experimental powder XRD pattern of Ce$_{2}$Ir$_{3}$Ga$_{5}$ (yellow open circles) and the Rietveld refinement (blue solid line). Small red bars show the Bragg reflection positions, and a green solid line (bottom) shows the difference between the calculated and observed intensities. (b) Inset: Partial crystal structure showing the atomic arrangement around the Ce site (green sphere).}
\end{figure}

\begin{table}[h!]
\caption{Structure parameters of Ce$_{2}$Ir$_{3}$Ga$_{5}$ obtained from powder XRD}
\begin{tabular}{l r} \hline\hline
Structure& Orthorhombic\\
Space group&        \textit{Ibam}\\ [1ex]
Lattice parameters\\ [0.5ex]
a (\text{\AA})&  10.0594(2)\\
b (\text{\AA})&  12.7239(3)\\
c (\text{\AA})&  5.7745(1)\\
V$_{Cell}$ (\text{\AA}$^{3}$)& 739.15(4)
\end{tabular}
\\[1ex]

\begingroup
\setlength{\tabcolsep}{3pt}
\begin{tabular}[b]{c c c c c c}
Atom&  Wyckoff position& x/a& y/b& z/c& U$_{iso/eq}$\\[1ex]
\hline
Ce1& 8j& 0.27267(5)& 0.36125(4)& 0& 86(1)\\
Ir1& 8j& 0.09465(4)& 0.14908(3)& 0& 88(1)\\
Ir2& 4b& 1/2& 0& 1/4& 75(1)\\                         
Ga1& 8j& 0.3274(1)& 0.0882(1)& 0& 89(2)\\
Ga2& 8g& 0& 0.2943(1)& 1/4& 85(2)\\
Ga3& 4a& 0& 0& 1/4& 92(3)\\
[1ex]
\hline
\end{tabular}
\par\medskip\footnotesize
\endgroup
\end{table}

Single-crystal XRD experiment was performed for crystal structure solution and refinement. A crystal~of regular shape was fixed with glue at the top of a glass needle for X-ray diffraction intensity data collection. According to reflection conditions and isotypism to U$_{2}$Co$_{3}$Si$_{5}$ type \cite{XR1,XR2}, the crystal structure of Ce$_{2}$Ir$_{3}$Ga$_{5}$ was solved by direct methods using SHELXS \cite{XR3} and refined using SHELXL \cite{XR3} in the orthorhombic space group \textit{Ibam} (No. 72). Further results obtained from the analysis of crystallographic data are provided in Table \ref{XRD1}, \ref{XRD2}, and \ref{XRD3} (See the supplemental file).
Powder X-ray diffraction (PXRD) intensities were collected for lattice parameter determination and phase purity analysis of Ce$_{2}$Ir$_{3}$Ga$_{5}$ bulk samples. Peak calibration followed by angle correction were performed on the PXRD pattern using the program WinXpow \cite{XR4} and taking LaB$_{6}$ (a = 4.1569 \AA) as an external standard. The lattice parameters a = 10.0594(2) \text{\AA}, b = 12.7239(3) \text{\AA}, c = 5.7745(1) \text{\AA} were determined by least-squares refinement on 169 reflection positions in the range of 5$\degree$ < 2$\theta$ < 118$\degree$ using the software WinCSD \cite{XR5}. Rietveld refinement on PXRD data (\figref{Fig1:xrd}) was performed using the Fullprof software \cite{XR6} by considering the crystallographic data obtained from the single crystal structure refinement. This result confirms quantitatively single-phase samples. The EDXS analysis, on polished pieces from both batches, results in good agreement (Ce$_{20.9}$Ir$_{29.7}$Ga$_{49.2}$) with the stoichiometric composition of Ce$_{20}$Ir$_{30}$Ga$_{50}$ obtained from the crystal structure refinement.
The crystal structure of the title compound (\figref{Fig1:xrd} Inset) shows two important features: i) a single cerium site and ii) the absence of a local inversion center at the Ce site.

\subsection{Magnetization}

\begin{figure} 
\includegraphics[width=1.0\columnwidth]{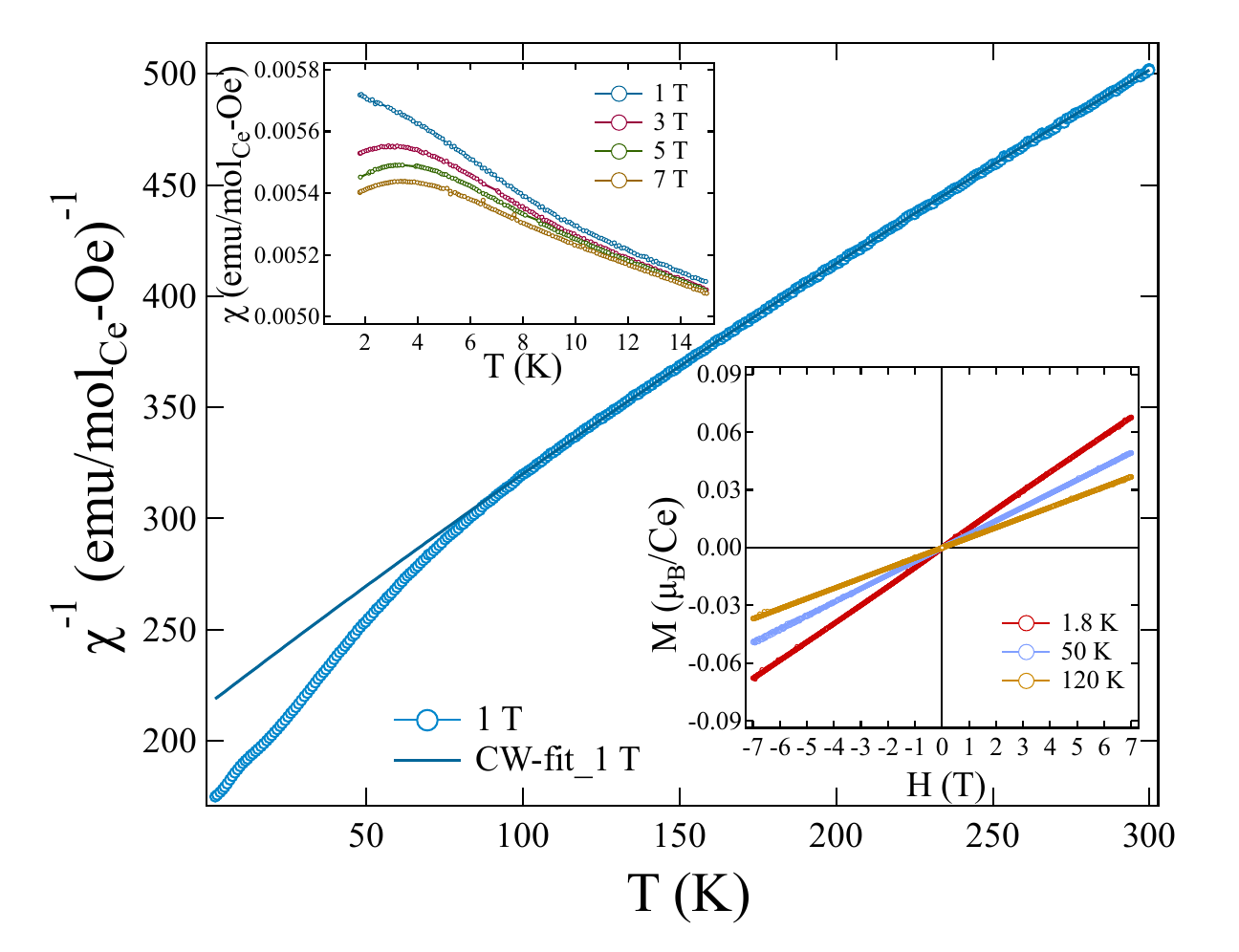} 
\caption{\label{Fig2:mag} Inverse susceptibility w.r.t. temperature in 1 T applied field is well described by a Curie-Weiss law in the high T regime. The inset on the top left represents $\chi$(T) at different applied fields, whereas the inset on the bottom right displays MH curves measured at different temperatures 1.8~K, 50 K, and 120 K.}
\end{figure}

Magnetization measurements have been performed in the zero-field cooled mode. The temperature dependence of the inverse of susceptibility $\chi^{-1}$(T) in 1 T external magnetic field is shown in \figref{Fig2:mag}. The high-temperature part (100 K $\leq$ T $\leq$ 300 K) of $\chi^{-1}$(T) shows a linear T dependence as expected for a local moment system. It can be well fitted with modified a Curie-Weiss equation: $\chi^{-1}$(T) = $\left(\chi_{0}+\frac{C}{T-\theta_{p}}\right)^{-1}$, where $\chi_{0}$ is a temperature independent contribution to the susceptibility, C is the Curie constant and $\theta_{p}$ is the Curie-Weiss temperature. The fit yields a small $\chi_{0}$ = 0.00038 emu/mol, C = 0.78(1) emu-mol$_{Ce}$/Oe, and $\theta_{p}$ = -184 (1) K. The effective moment deduced from C is $\mu_{eff}$ = 2.5$\mu_{B}$, which is close to the expected free ion value (2.54$\mu_{B}$) for Ce$^{3+}$. Non-linearity for T < 100 K is caused by crystal electric field (CEF) effects, i.e., a depopulation of excited CEF levels. There is no clear anomaly in the T dependence of the susceptibility, indicating the absence of a well-defined magnetic order. At small fields, there is a tiny kink in $\chi$(T) at around 10 K. Because this anomaly is quite small, visible only at low fields, and there is no corresponding anomaly in other properties like resistivity or specific heat, we suspect this kink might originate from a tiny amount of disordered ferromagnetic foreign phase, which remained undetected in the EDXS measurements. While at low fields below 2 T, $\chi$(T) continuously increases down to the lowest investigated temperatures, application of higher fields results in the appearance of a maximum in $\chi$(T) at around 4 K (Upper inset of \figref{Fig2:mag}). Such a maximum in $\chi$(T) has commonly been observed in Ce-based Kondo lattice systems located slightly on the non-magnetic side of the QCP, like, e.g., CeRu$_{2}$Si$_{2}$ \cite{CeRu2Si2} or CeNi$_{2}$Ge$_{2}$ \cite{CeNi2Ge2_1,CeNi2Ge2_2}. M(H) isotherms taken at different temperatures, 1.8 K, 50 K, and 120 K (lower inset of \figref{Fig2:mag}), all exhibit a linear increase with increasing field without sign of saturation up to the highest magnetic fields. In summary, the susceptibility results indicate a localized trivalent Ce state at higher temperatures without clear evidence for a magnetic ordering at low T.

\subsection{Electrical Resistivity}

\begin{figure}
\includegraphics[width=1.0\columnwidth,origin=b]{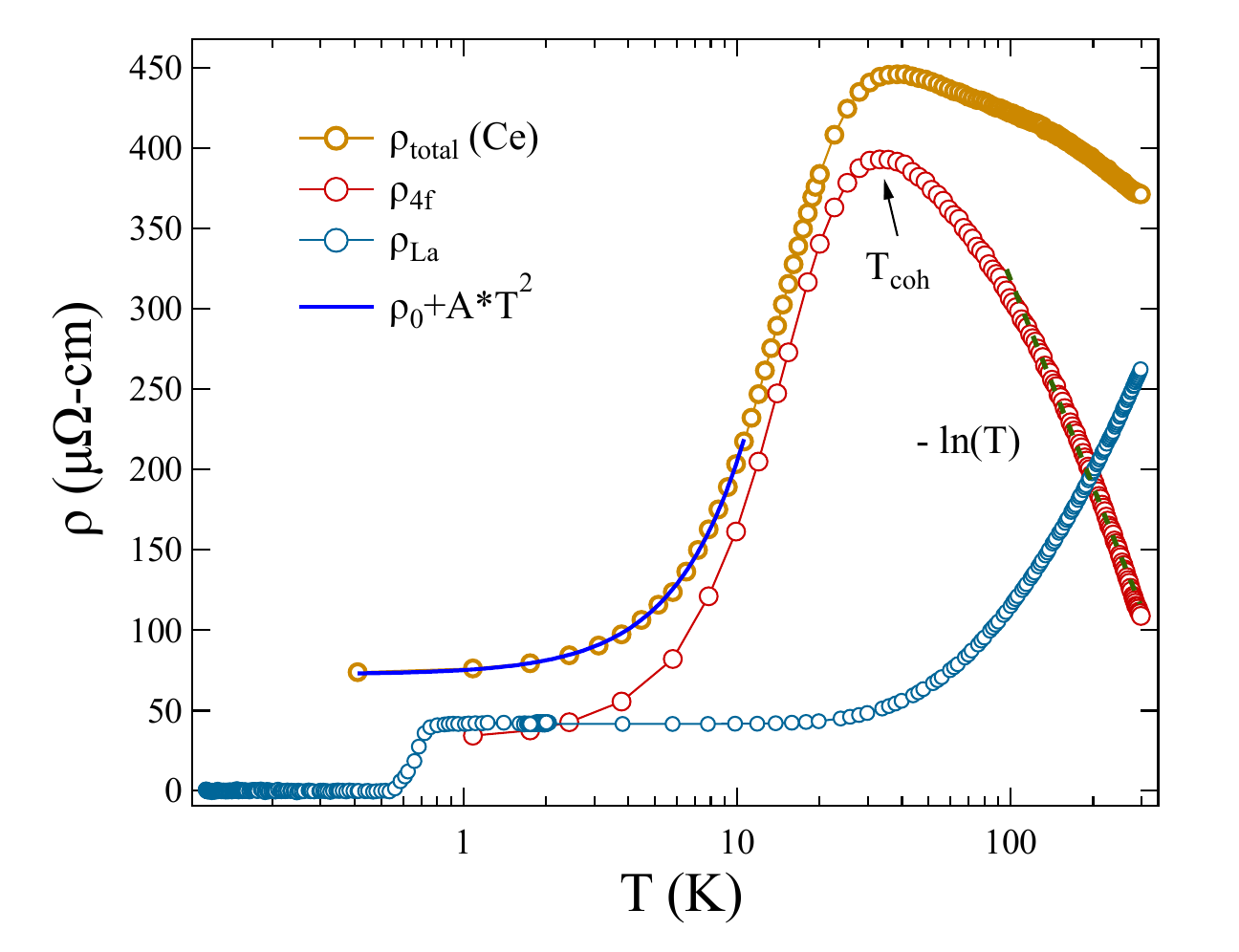}
\caption{\label{Fig3:rho} Temperature dependence of electrical resistivity for (Ce,La)$_{2}$Ir$_{3}$Ga$_{5}$ in zero applied fields where T is on a logarithmic scale. The total contribution to resistivity is shown by brown open circles, whereas the magnetic contribution is represented via red open circles.}
\end{figure}

The temperature dependence of the electrical resistivity $\rho$(T) of both Ce$_{2}$Ir$_{3}$Ga$_{5}$ and its non-f analog La$_{2}$Ir$_{3}$Ga$_{5}$ are shown in \figref{Fig3:rho}. The La-based compound displays the typical behavior of a standard metal, with a constant residual resistivity at low T and a linear increase due to phonon scattering at high T. Interestingly, it also shows a broadened transition to a superconducting state with an onset T$_{c}$ and an offset T$_{c}$ at about 0.75 K and 0.55 K, respectively. In contrast, $\rho$(T) of Ce$_{2}$Ir$_{3}$Ga$_{5}$ is completely different, with an increase with decreasing T in the range 50 K < T < 300 K, a broad maximum at around 30 K, followed by a strong decrease to lower T and a T$^{2}$ dependence below 10 K. This kind of temperature dependence is typical for a Kondo lattice system close to the QCP, but on the non-magnetic side of the QCP. Since all these features are connected to the magnetic scattering of the 4$f$ electrons, we tentatively estimated this magnetic contribution by subtracting $\rho$(T) of La$_{2}$Ir$_{3}$Ga$_{5}$, making those features more apparent, especially at high T. There is an increase of $\rho_{mag}$ with decreasing temperatures in the high-temperature region following a logarithmic behavior as expected for an incoherent Kondo-type scattering. The maximum at around 30 K indicates the onset of coherence between the individual Kondo sites and is thus denominated by T$_{coh}$. Usually, it provides a first estimation of the Kondo energy scale, the Kondo temperature T$_{K}$. The slight bump in $\rho_{mag}$(T) at about 150 K is likely connected with the Kondo scattering of excited CEF levels, suggesting the first excited CEF doublet to be in the range 150 K - 300 K, a typical value for Cerium based intermetallics. The strong decrease of $\rho_{mag}$ below 20 K is due to the formation of a coherent Kondo state where the scattering of the Kondo sites gets in phase and thus becomes elastic. The T$^{2}$ dependence at low temperatures reflects the formation of a Fermi liquid. A fit with $\rho$(T) = $\rho_{0}$ + A$\cross$T$^{n}$ gives $\rho_{0}$ = 72.5 $\mu \Omega$-cm, A = 2.7 $\mu \Omega$-cm K$^{-2}$, and n = 1.7 $\approx$ 2 below 10 K. However, we are not confident about the absolute value of A and $\rho_{0}$. The reason is that the absolute values of $\rho$(T) at high T are larger than expected for such metallic systems\cite{CeNiAl4,CeCu9In2}. We suspect this might be due to the formation of cracks within the sample, which reduces the effective width of the conduction channel and thus results in apparently high $\rho$(T) values. However, such problems should not affect the resistivity ratios. Residual resistivity ratio (RRR) of 5 and 6 for Ce$_{2}$Ir$_{3}$Ga$_{5}$ and La$_{2}$Ir$_{3}$Ga$_{5}$ respectively, are typical for this kind of systems. They indicate reasonably pure compounds with a small amount of disorder.

\subsection{Specific Heat}

\begin{figure}
\includegraphics[width=1.0\columnwidth,origin=b]{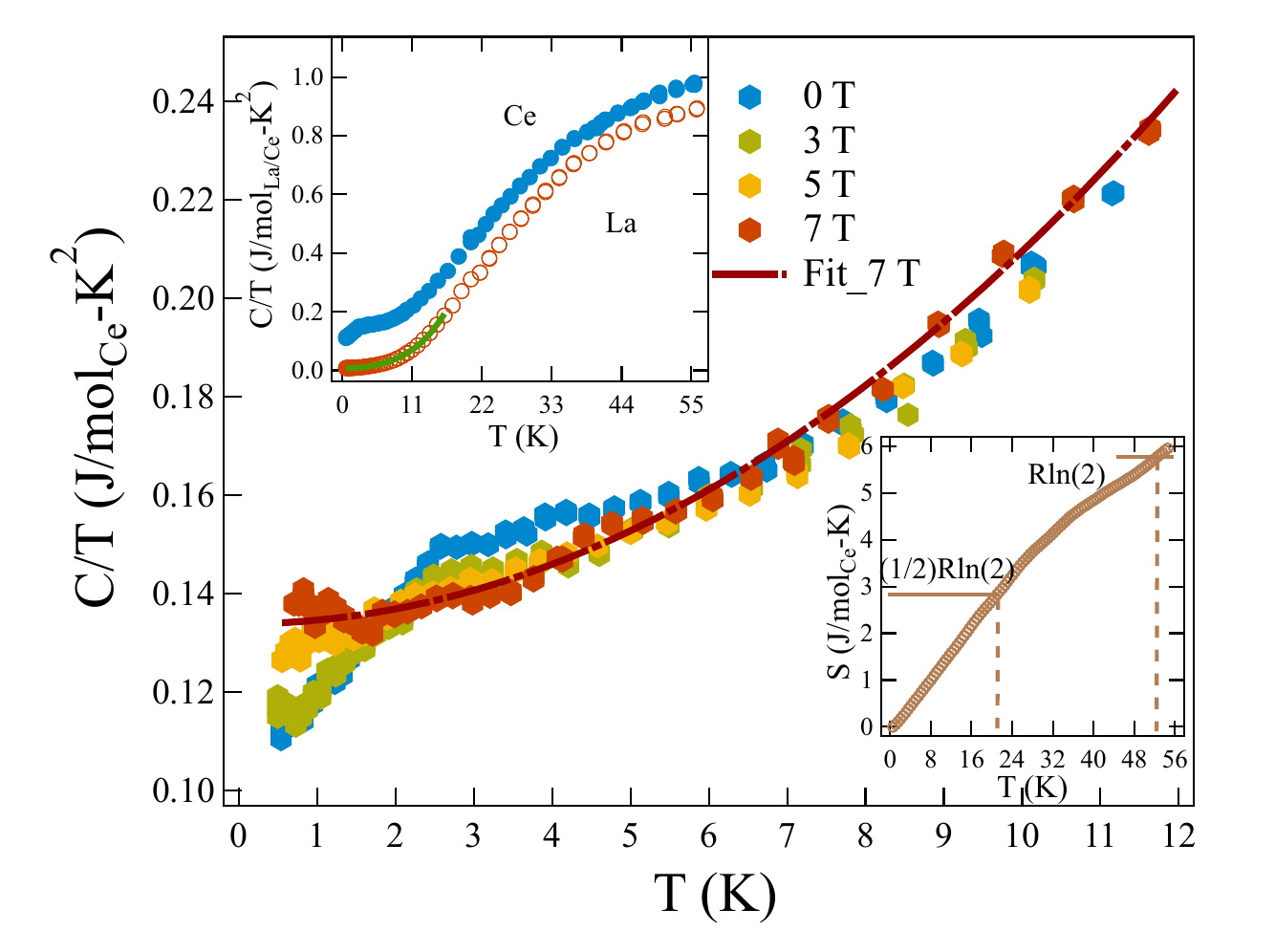}
\caption{\label{Fig4:SH} C/T as a function of temperature in zero and different applied fields for Ce$_{2}$Ir$_{3}$Ga$_{5}$. The dashed line represents a fit to 7 T field data with C/T = $\gamma$+$\beta$T$^{2}$. The lower inset shows the calculated entropy, whereas the upper inset displays the specific heat of (Ce,La)$_{2}$Ir$_{3}$Ga$_{5}$ in zero field.}
\end{figure}

The upper inset in \figref{Fig4:SH} displays the temperature dependence of specific heat $C$, normalised by temperature, C(T)/T, for Ce$_{2}$Ir$_{3}$Ga$_{5}$ and its non-4$f$ analog La$_{2}$Ir$_{3}$Ga$_{5}$ in zero applied fields. La$_{2}$Ir$_{3}$Ga$_{5}$ displays a standard metallic behavior where the high-temperature part is accounted by the phononic contribution and the lowest temperature region, which is nearly constant below 1.8 K, provides the electronic contribution. A fit to the data in the range 0.5 K $<$ T $<$ 16 K using C/T = $\gamma$ + $\beta_{3} T^{2}$ + $\beta_{5} T^{4}$ where $\gamma$, $\beta_{3}$, and $\beta_{5}$ are the electronic and phononic contributions, respectively is shown by green solid line. It provides $\gamma$ = 8.1 mJ(mol$_{La}$)$^{-1}$K$^{-2}$, a moderate value for metallic compounds, $\beta_{3}$ = 0.34 mJ(mol$_{La}$)$^{-1}$K$^{-4}$, and $\beta_{5}$ = 1.44 $\mu$J(mol$_{La}$)$^{-1}$K$^{-6}$. By using the $\beta_{3}$ value, Debye temperature was calculated from the expression: $\theta_{D}= \left(\frac{12\pi^{4}RN}{5\beta_{3}}\right)^{\frac{1}{3}}$ = 306 K, also a quite typical value for such a compound. There N = 5, the number of atoms per formula unit according to per La mol, and R is the gas constant. The parameter $\beta_{5}$ accounts for the deviation of the phonon spectra from simple Debye spectra. The specific heat of La$_{2}$Ir$_{3}$Ga$_{5}$ did not exhibit a superconducting transition to the lowest measured temperature (0.53 K), in contrast to $\rho$(T) data. We suspect that the low T limit achieved in specific heat measurement, T = 0.53 K, was not sufficiently low since usually the onset of the SC transition in C(T) is only seen at the offset ($\rho$ = 0) of the transition in $\rho$(T).

In comparison, the specific heat of Ce$_{2}$Ir$_{3}$Ga$_{5}$ is larger than that of La$_{2}$Ir$_{3}$Ga$_{5}$ in the whole T range. The difference C$_{4f}$(T) = C$_{Ce}$(T) - C$_{La}$(T) can be attributed to the effect of the 4$f$ electrons. At first glance, it results in C$_{4f}$(T)/T being only weakly T dependent in the investigated T range, reminiscent of a Fermi liquid behavior. This indicates that the 4$f$ electrons do not behave as localized magnetic moments but instead hybridize strongly with the conduction electrons, forming heavy quasi-particles. Below 10 K, the C/T values are of the order of 150 mJ(mol$_{Ce}$)$^{-1}$K$^{-2}$, thus enhanced by a factor of 20 compared to the non-4$f$ analog La$_{2}$Ir$_{3}$Ga$_{5}$. It provides evidence that Ce$_{2}$Ir$_{3}$Ga$_{5}$ is a moderately heavy fermion system with strong f-c hybridization. By integrating C$_{4f}$(T)/T we estimated the entropy S$_{4f}$(T) connected with the 4$f$ electrons (lower inset in \figref{Fig4:SH}). From the J = 5/2 multiplet for a localized Ce-4$f$ electron, one expects a high-temperature limit S$_{4f}$($\infty$) = Rln6, but since the crystal electric field (CEF) splits this multiplet into three doublets, usually only the contribution Rln2 of the ground state doublet is recovered at lower T. In Ce$_{2}$Ir$_{3}$Ga$_{5}$, the 4$f$ entropy increases continuously up to the highest investigated T, with only a small decrease in the slope above 35 K, and reaches Rln2 at about 52~K. This almost linear increase and the absence of saturation at Rln2 suggest that the energy scale of the f-c hybridization is similar to the CEF scale, resulting in a broadening and mixing of all CEF levels. The temperature T(0.5Rln2) = 21 K at which half of Rln2 is reached frequently provides a good estimate of the Kondo scale, with T$_{K}$ = 2T(0.5Rln2). This results in T$_{K}$ = 42 K, which is in reasonable agreement with the temperature of the maximum in $\rho$(T). Thus, both the resistivity and the specific heat indicate Ce$_{2}$Ir$_{3}$Ga$_{5}$ to bear strong $f$-$c$ hybridization resulting in a moderately heavy fermion system with a Kondo scale of the order of 40 K.

A closer look at the specific heat data at low T reveals a small kink in C(T)/T at T$_{0}$ $\approx$ 2.7 K, below which C(T)/T decreases with T, suggesting the presence of a phase transition at T$_{0}$. The obvious question is whether this small anomaly is an intrinsic property of Ce$_{2}$Ir$_{3}$Ga$_{5}$, or due to a tiny amount of undetected foreign phase. The small size of this anomaly and the very small amount of entropy connected with this transition, only about 4\% Rln2, favors the latter scenario. However, an analysis of the field dependence of C/T provides arguments in favor of the intrinsic nature of this transition. Application of a magnetic field H $>$ 3 T weakens this anomaly, seemingly without shifting T$_{0}$ to lower T, and at 7 T, it can no longer be resolved (main part of \figref{Fig4:SH}). The suppression occurs in such a way that for T $<$ 0.5T$_{0}$ the specific heat gets enhanced. At the same time, it gets suppressed above 0.5T$_{0}$. A check of the entropy S(T, B) confirms that the entropy at 10 K is conserved, i.e., S(10 K) is independent of the magnetic field. That means that the field-induced suppression of this transition is connected with a transfer of entropy gain from high T $>$ 0.5T$_{0}$ in the ordered regime to low T $<$ 0.5T$_{0}$ in the high field regime. For magnetic order of a local moment magnetic system, one expects and usually observes precisely the opposite. In contrast, such a transfer of entropy to lower T is reminiscent of phase transitions in systems of itinerant electrons, e.g., a CDW or a SC transition. The difference in C/T at the lowest T between the 7 T and the zero field curves amounts to about 20\% of the C/T value at 7 T, $\gamma_{7T}$ = 134 mJ(molCe)$^{-1}$K$^{-2}$. Assuming the transition to be intrinsic to Ce$_{2}$Ir$_{3}$Ga$_{5}$, that would imply 20\% of the density of states (DOS) of the heavy quasi-particles gets condensed, an amount quite typical, e.g. for a charge density wave (CDW) transition. Assuming the transition to be due to a foreign phase, it would necessitate two conditions: firstly, that a substantial portion of the observed C/T at low T, exceeding 30\%, is due to this foreign phase, and secondly, that a huge percentage of the DOS of this foreign phase gets condensed at T$_{0}$. Both seem unlikely. Assuming the transition to be intrinsic to Ce$_{2}$Ir$_{3}$Ga$_{5}$, the very small size of the related anomaly and the condensed entropy would be in perfect accordance with the ordering temperature T$_{0}$ being one order of magnitude smaller than the Kondo temperature. We note that the closely related compound Ce$_{2}$Ru$_{3}$Ge$_{5}$ \cite{Ce2Ru3Ge5}, which crystallizes in the same structure type, presents a ferromagnetic heavy fermion ground state with T$_{C}$ = 7.9 K and an estimated T$_{K}$ $\approx$ 19 K. Because of the much larger T$_{C}$/T$_{K}$ ratio, the size of the anomalies and the related entropy gain is much larger in Ce$_{2}$Ru$_{3}$Ge$_{5}$. From the total count of valence electrons Ce$_{2}$Ru$_{3}$Ge$_{5}$ is close to Ce$_{2}$Ir$_{3}$Ga$_{5}$, since Ru has one valence electron less than Ir, but Ga has one more. Thus, Ce$_{2}$Ir$_{3}$Ga$_{5}$ could be a homolog of Ce$_{2}$Ru$_{3}$Ge$_{5}$ pushed very close to the critical point where magnetic order disappears. However, as the evidence remains inconclusive, it is currently not feasible to definitively determine the nature of this transition, whether it is intrinsic or not.

\section{Summary}
We explored the Ce-Ir-Ga ternary system, in the search for a new cerium-based system that presents a locally non-centrosymmetric environment with preserved global inversion symmetry and has an appropriate strength of $f$-$c$ hybridization so that an unconventional superconducting state like CeRh$_{2}$As$_{2}$ can be realized. There, the combination of Ir and Ga can provide the fine-tuned hybridization strength. This search leads to the discovery of a new compound Ce$_{2}$Ir$_{3}$Ga$_{5}$. Structural characterization revealed an orthorhombic U$_{2}$Co$_{3}$Si$_{5}$ structure type with space group: I$bam$, No. 72. Magnetization measurements suggest a localized Ce$^{3+}$ valence state at high-temperatures but no clear sign of magnetic ordering down to the lowest temperature. In AC transport measurements, $\rho$(T) displays a typical Kondo behavior with a logarithmic increase in resistivity at high temperatures and a pronounced decrease below T$_{coh}$ $\approx$ 30 K. Below 10 K, $\rho$(T) follows a T$^{2}$ dependence, consistent with a Fermi liquid picture. Specific heat measurements evidence an enhanced electronic contribution with an almost constant C/T value, $\gamma$ = 150 mJ(molCe)$^{-1}$K$^{-2}$ in the range 3 K < 30 K. This indicates strongly renormalized quasi-particles induced by a sizeable f-c hybridization. From the T dependence of the entropy, we estimated a Kondo scale of the order of 42 K, which is in agreement with the temperature of the maximum in $\rho$(T). Thus, resistivity and specific heat measurements indicate Ce$_{2}$Ir$_{3}$Ga$_{5}$ as a moderately heavy Fermion system with a reasonably large Kondo scale. Susceptibility and resistivity measurements do not present any anomaly related to a transition, suggesting a paramagnetic heavy Fermi liquid ground state. However, specific heat data reveal a small anomaly at T$_{0}$ = 2.7 K. At first glance, one would suspect this anomaly is due to a foreign phase. However, a more detailed analysis suggests that it could also be due to an intrinsic itinerant type of order. Therefore, at present, we cannot conclude whether Ce$_{2}$Ir$_{3}$Ga$_{5}$ is on the magnetically ordered or the non-ordered side of the quantum critical point. Further studies such as $\mu$SR and NMR experiments are needed. In addition, it will be interesting to observe whether further improvements in the sample quality result in the onset of unconventional superconductivity.

\section{Acknowledgments}

We thank S. Kostmann and P. Scheppan for EDXS measurements, as well as S. Scharsach and M. Schmidt for performing DTA experiment.

\clearpage

\title{Supplementary Information to "Ce$_{2}$Ir$_{3}$Ga$_{5}$: a new locally non-centrosymmetric heavy fermion system"}\label{Struc_para}
\maketitle

\subsection*{Structural Characterization}\label{XRD1}
Table \ref{XRD1} presents the details of the crystallographic data and handling of single crystal data collection, while Table \ref{XRD2} lists the refined atomic coordinates and atomic displacement parameters. Selected interatomic distances are provided in Table \ref{XRD3}. A X-ray powder diffraction pattern was calculated using the atomic parameters obtained from the single crystal data, employing the WinXPow software. This calculation yielded a good agreement with the experimental data as shown in the main \figref{Fig1:xrd}.
\setcounter{table}{0}
\renewcommand{\thetable}{S\arabic{table}}
\begin{table}[H]
\caption{Crystallographic data for Ce$_{2}$Ir$_{3}$Ga$_{5}$. Lattice parameter obtained from powder XRD data, chemical composition from EDXS analysis.}
\label{XRD1}
\begingroup
\setlength{\tabcolsep}{1pt}
\begin{tabular}{l r} \hline\hline
Composition& Ce$_{2}$Ir$_{3}$Ga$_{5}$\\
Molar mass&       1205.50\\ [0.5ex]
Crystal color, shape& Gray, prismatic\\ [0.5ex]
Crystal dimensions (mm$^{3}$)& 0.032$\times$0.038$\times$0.042\\[0.5ex]
Space group, Z&	$Ibam$ (No. 72), 4\\[0.5ex]
Lattice parameters (\text{\AA})& a = 10.0594(2)\\[0.5ex]
T = 293 K& 	b = 12.7239(3)\\[0.5ex]
&  c =   5.7745(1)\\[0.5ex]
V (10$^{6}$ pm$^{3}$), $\rho$ (g cm$^{-3}$)&	739.11(5), 10.832(1) \\[0.5ex]
Diffractometer, detector&	Rigaku AFC7. CCD, \\[0.5ex]
&  Saturn724+ \\[0.5ex]
Radiation&	Mo K$_{\alpha}$ ($\lambda$ = 0.71073 \text{\AA})\\[0.5ex]
Exposures, steps&	1200, $\phi$ = 0.6$\degree$\\[0.5ex]
Absorption correction&	Multi-scan ($\mu$ = 83.61 mm$^{-1}$)\\[0.5ex]
T$_{min}$/T$_{max}$&	0.030/0.069\\[0.5ex]
2$\theta_{max}$; sin$\theta$/$\lambda$&	71.29$\degree$; 0.82\\[0.5ex]
$hkl$ range&	-11 < $h$ < 16\\[0.5ex]
&-20 < $k$ < 20\\[0.5ex]
&-6 < $l$ < 9\\[0.5ex]
Reflections:& \\[0.5ex]
Measured& 6744 \\[0.5ex]
Used in refinement& 878	\\[0.5ex]
R$_{(eq)}$; R$_{(sigma)}$&	0.033; 0.020 \\[0.5ex]
Observation criteria&	F($hkl$) > 4$\sigma$ F($hkl$) \\[0.5ex]
Refinement&	Full-matrix least-squares on F$^{2}$ \\[0.5ex]
Parameters&	31 \\[0.5ex]
R(F), R$_{w}$& 0.029, 0.056 \\[0.5ex]
Goodness of fit& 1.167 \\[0.5ex]	
$\Delta \rho_{min}$, $\Delta \rho_{max}$ (e $\text{\AA}^{-3}$) &	-4.62, 4.64 \\[0.5ex] \hline\hline
\end{tabular}
\endgroup
\end{table}

\renewcommand{\thetable}{S\arabic{table}}
\begin{table}[H]
\caption{Atomic positions and displacement parameters [pm$^{2}$] for Ce$_{2}$Ir$_{3}$Ga$_{5}$ (U$_{eq}$ = 4/3[U$_{11}$ a*$^2$ a$^2$ + ... 2 U$_{23}$ b* c* b c cos$\alpha$])}
\label{XRD2}
\begingroup
\setlength{\tabcolsep}{1pt}
\begin{tabular}[b]{c c c c c c c}\hline
$\bold{Atom}$&  $\bold{Ce}$& $\bold{Ir1}$& $\bold{Ir2}$& $\bold{Ga1}$& $\bold{Ga2}$& $\bold{Ga3}$ \\[1ex]
\hline
Site&	8$j$&	8$j$&	4$b$&	8$j$&	8$g$&	4$a$ \\[0.5ex]
x/a&	0.27267(5)&	0.09465(4)&	1/2&	0.3274(1)&	0&	0 \\[0.5ex]
y/b&	0.36125(4)&	0.14908(3)&	0&	0.0882(1)&	0.2943(1)&	0 \\[0.5ex]
z/c&	0&	0&	1/4&	0&	1/4&	1/4 \\[0.5ex]
U$_{iso/eq}$&	86(1)&	88(1)&	75(1)&	89(2)&	85(2)&	92(3) \\[0.5ex]
U$_{11}$&	75(2)&	81(2)&	83(2)&	70(4)&	83(4)&	113(6) \\[0.5ex]
U$_{22}$&	109(2)&	83(2)&	83(2)&	107(4)&	93(4)&	75(6) \\[0.5ex]
U$_{33}$&	75(2)&	99(2)&	58(2)&	90(4)&	80(4)&	86(6) \\[0.5ex]
U$_{12}$&	0(2)&	4(1)&	0&	5(3)&	0&	0 \\[0.5ex]
U$_{13}$&	0&	0&	U$_{12}$&	0&	5(4)&	U$_{12}$ \\[0.5ex]
U$_{23}$&	U$_{13}$&	U$_{12}$&	U$_{12}$&	U$_{13}$&	U$_{12}$&	U$_{12}$ \\[0.5ex]
\hline
\end{tabular}
\par\medskip\footnotesize
\endgroup
\end{table}

\renewcommand{\thetable}{S\arabic{table}}
\begin{table}[H]
\caption{Interatomic distances [\text{\AA}] in Ce$_{2}$Ir$_{3}$Ga$_{5}$}
\label{XRD3}
\begingroup
\setlength{\tabcolsep}{6pt}
\begin{tabular}[b]{c l l c l l}\hline
$\bold{Ce}$&	1$\times$Ga1&	3.058(1)&	$\bold{Ga1}$&	1$\times$Ir1&	2.466(1)\\
	& 2$\times$Ga1&	3.125(1)&		    &2$\times$Ir2&	2.521(1)\\
	& 1$\times$Ir1&	3.1835(3)&		    &2$\times$Ga2&	2.708(1)\\ 
	& 2$\times$Ga2&	3.214(1)&		    &1$\times$Ce&	   3.058(1)\\
	& 2$\times$Ga3&	3.230(1)&		    &2$\times$Ce&	   3.1247(5)\\
	& 2$\times$Ir1&	3.240(1)&		   &1$\times$Ce&	3.518(1)\\
	& 1$\times$Ir1&	3.242(1)&	       &           &              \\
	& 1$\times$Ga2&	3.351(1)&	$\bold{Ga2}$& 2$\times$Ir1&	2.531(1)\\
	& &	&	&	1$\times$Ir2&	2.617(1)\\
 	&       &     &    &			2$\times$Ga1&	2.708(1)\\
$\bold{Ir1}$&	1$\times$Ga1&	2.466(1)&		&2$\times$Ga2&	 2.8873(1)\\
	&1$\times$Ga2&	2.531(1)&	&	1$\times$Ce&	3.214(1)\\
	&1$\times$Ga3&	2.5669(3)&	&	1$\times$Ce&	3.351(1)\\
	&2$\times$Ce&	3.1835(3)&	&	&	\\
	&2$\times$Ce&	3.239(1)&	$\bold{Ga3}$&	3$\times$Ir1&	2.5668(3)\\
	&1$\times$Ce&	3.242(1)&                 &1$\times$Ir1&	2.6669(3)\\
	& &	&		                              &2$\times$Ga3&	2.8873(1)\\    
                                  
$\bold{Ir2}$&	4$\times$Ga1&	2.521(1)&		&4$\times$Ce&	3.230(1)\\
	&2$\times$Ga2&	2.618(1)&		&&\\
	&2$\times$Ir2&	2.8875(1)&		&&\\
	&4$\times$Ce&	3.567(1)\\[0.5ex]		\hline

\hline
\end{tabular}
\par\medskip\footnotesize
\endgroup
\end{table}

\end{document}